# Electronic Dispersion Pre-Compensation in PM-QPSK Systems over Mixed-Fiber Links

A. Carena(1), Y. Jiang(1), P. Poggiolini(1), G. Bosco(1), V. Curri(1), F. Forghieri(2)

(1) Politecnico di Torino, DET, corso Duca degli Abruzzi, 24,10129 Torino, Italy, carena@polito.it
(2) Cisco Photonics Italy srl, via Philips 12, 20900 Monza, fforghie@cisco.com

**Abstract** *Dispersion pre-compensation is shown to potentially lead to a substantial non-linearity reduction in PM-QPSK links that use a mixture of high and low dispersion fibers. However, the much larger PAPR of the pre-compensated signal poses challenging requirements on the transmitter.*

## Introduction

Chromatic dispersion pre-compensation (CDP) was extensively used and provided substantial performance gains in IMDD dispersion-managed systems. After the advent of coherent systems operating over uncompensated-transmission (UT) links, several studies have been carried out to find out whether CDP would be useful in this new system scenario as well, both simulative [1],[2] and experimental [3]-[6]. It was found that the optimum CDP, over links with same fiber and span length ("homogenous links"), amounts to 50% of the total link accumulated dispersion. The potential gain on either the Q factor or the maximum reach was however modest, on the order of a fraction of a dB. The highest gain was 0.9 dB on $Q^2$ in [4], where PM-BSPK was used.

Inhomogenous links have been investigated, too. In [6], a mixed-fiber link was addressed where a multi-span section used one fiber type and another section a different type, with the two types being either SMF or TrueWave RS. The optimum CDP was for zero-accumulated-dispersion being at the middle of the lower-dispersion and more non-linear Truewave RS fiber section. The $Q^2$ gain was however minimal, about 0.2 dB at the optimum launch power.

Most experiments partly blamed the low gains on penalties incurred at the transmitter (Tx) when imposing CDP. This is because CDP greatly increases the signal peak-to-average power ratio (PAPR), which stresses limited-resolution DACs and the non-linear Mach-Zehnder modulator trans-characteristic. It is therefore conceivable that higher gains are possible, which were masked by implementation penalties. This is especially true for mixed-fiber links which are still largely unexplored, with only a few pioneering studies available, such as [6].

One of the problems in the investigation of CDP systems has also been the lack of models capable of accurately predicting their non-linear behavior. The GN model [7], for example, cannot deal with this situation because it assumes that the signal behaves as Gaussian noise (as if it had undergone "infinite" CDP). While this approximation makes the model very simple, it prevents it from being able to account for CDP. Recently, however, extending and generalizing a procedure proposed in [8], an enhanced GN model (EGN model) has been derived [9]. Its structure is similar to the GN model but it avoids making the signal-Gaussianity assumption, enabling it to account for CDP.

In this paper we propose a new detailed assessment of CDP over PM-QPSK UT links, both simulative and analytical. We use the EGN model to back up all simulative results. We found very good agreement between theory and simulations, which provided us with strong cross-confirmation of the results.

We first analyzed homogenous links. We found that indeed the advantage of CDP appears to be modest, about 0.1 to 0.15dB over $Q^2$, potentially increasing the maximum system reach by only a few percent. We then addressed a mixed-fiber case where SMF is followed by LS fiber. We show that in this case the potential $Q^2$ gain at optimum CDP and optimum launch power is non-negligible (0.76 dB). However, we point out that there could be substantial implementation hurdles that may thwart its actual fruition.

## PM-QPSK system set-up

The symbol rate ($R_s$) was set to 32 GBaud. The Tx generated pulses with a root-raised-cosine spectrum, with roll-off equal to 0.05. CDP was applied by digital filtering. Four ideal DACs generated the signals driving two nested Mach-Zehnder modulators, operated in their linear trans-characteristic range. The channel spacing was set to 33.6 GHz, i.e., the minimum spacing still ensuring no inter-channel crosstalk. The number of WDM channels was set to 15. Three fiber types were considered: SMF, NZDSF and LS. Their parameters were, respectively: dispersion 16.7, 3.8 and -1.8 ps/(nm·km); attenuation 0.2, 0.22 and 0.22 dB/km; non-linearity coefficient 1.3, 1.5 and 2.1 $(W\cdot km)^{-1}$. Lumped amplification through EDFAs was assumed, with noise figure 5 dB. The Rx used a customary optical front-end, feeding four balanced photo-detectors. They were followed by electrical anti-alias 5-pole Bessel filters with bandwidth $R_s/2$. Their output was sampled at 2 samples per symbol. Then, electronic CD post-compensation was performed, followed by an adaptive 2x2 equalizer, driven by a decision-directed LMS algorithm. Carrier and phase recovery were not needed since the Tx and local oscillator lasers were assumed ideal.

All test links were UT. For the simulations, data were generated using multiple independent PRBS's of length ($2^{16}$-1), four for each PM-QPSK channel. PRBSs were different from channel to channel. The total simulation length was $2^{17}$ symbols. The simulations were carried out over the whole 15-channel WDM band and used stringent accuracy settings (similar to [7]).

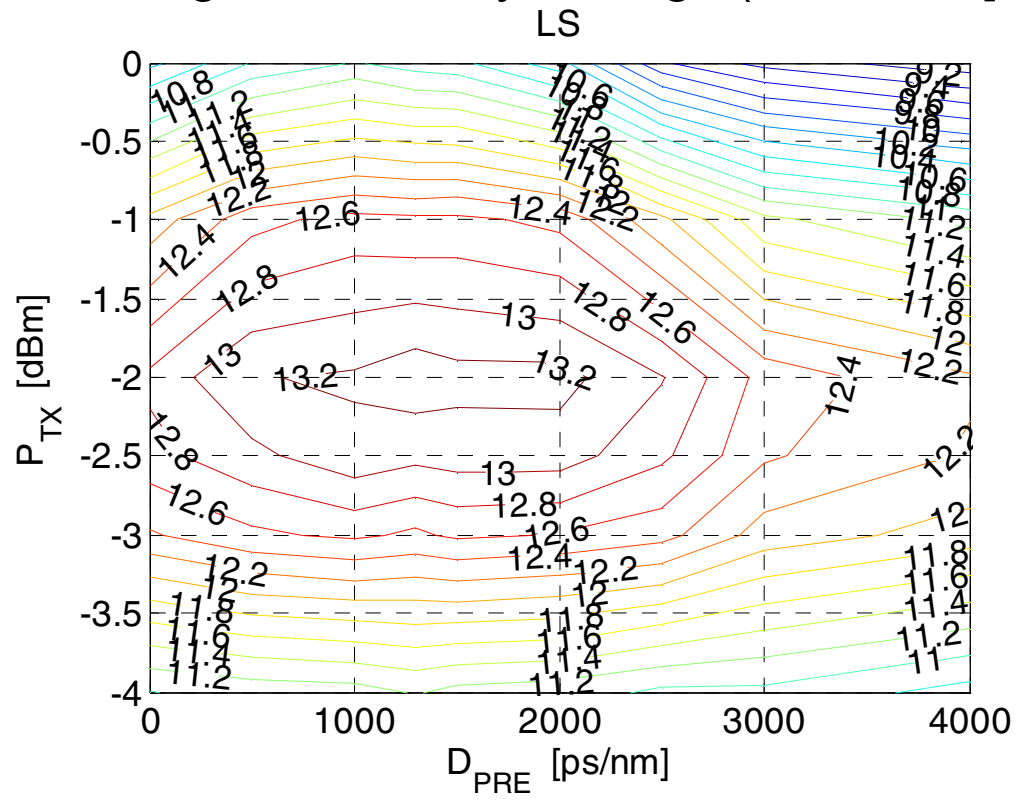

**Fig. 1:** System max-reach in number of spans (interpolated to non-integer for smoothness), vs. launch power per channel $P_{TX}$ and pre-compensation $D_{PRE}$. System is quasi-Nyquist WDM PM-QPSK, 15 channels, 32 GBaud, LS fiber.

## Homogeneous fiber links

The homogenous (single fiber type) link used 120km span length. We first characterized the system reach, for a target BER $10^{-2}$. Its value depends on both $P_{TX}$ and $D_{PRE}$. The results are displayed in Fig.1. We show the LS case only, for lack of space, but the same qualitative behavior is found for SMF and NZDSF, and all comments made for LS apply to the other fibers too. The plot shows the presence of a *maximum* (max-reach), achieved for $D_{PRE}$=1400 ps/nm, that is 50% of the link accumulated dispersion, confirming prior results [1],[2],[4],[5]. The optimum $P_{TX}$ appears to depend very weakly on $D_{PRE}$. The actual improvement in max-reach vs. no CDP ($D_{PRE}$=0) is however quite small. It is 3.6% for LS (see Fig.1) and about 3% for NZDSF and SMF. The respective max-reaches are 13.3, 19.4 and 43.7 spans.

To better understand the effect, we studied in detail the accumulation of non-linear interference (NLI) noise along the link. The estimation of the NLI noise power was performed as follows. We set $P_{TX}$ to the value granting max-reach ($P_{TXmax}$). The Rx electrical noise variance of each signal point of the constellation was evaluated on both quadratures and polarizations, turning off ASE noise. The variances were then averaged to obtain an estimate of the NLI power $P_{NLI}$ impinging on the Rx. We then calculated the quantity $SNR_{NLI}$ defined as:

$$SNR_{NLI} = P_{TXmax} / P_{NLI} \qquad (1)$$

In Fig. 2 we plot the $SNR_{NLI}$ results, vs. number of spans $N_{span}$ (SMF omitted for lack of space). The range of $N_{span}$ goes from 1 to the max-reach. The blue curves are for no CDP, the red curves are for optimum CDP. Dashed lines are analytical (EGN model), solid lines are simulations. The black line is the GN model.

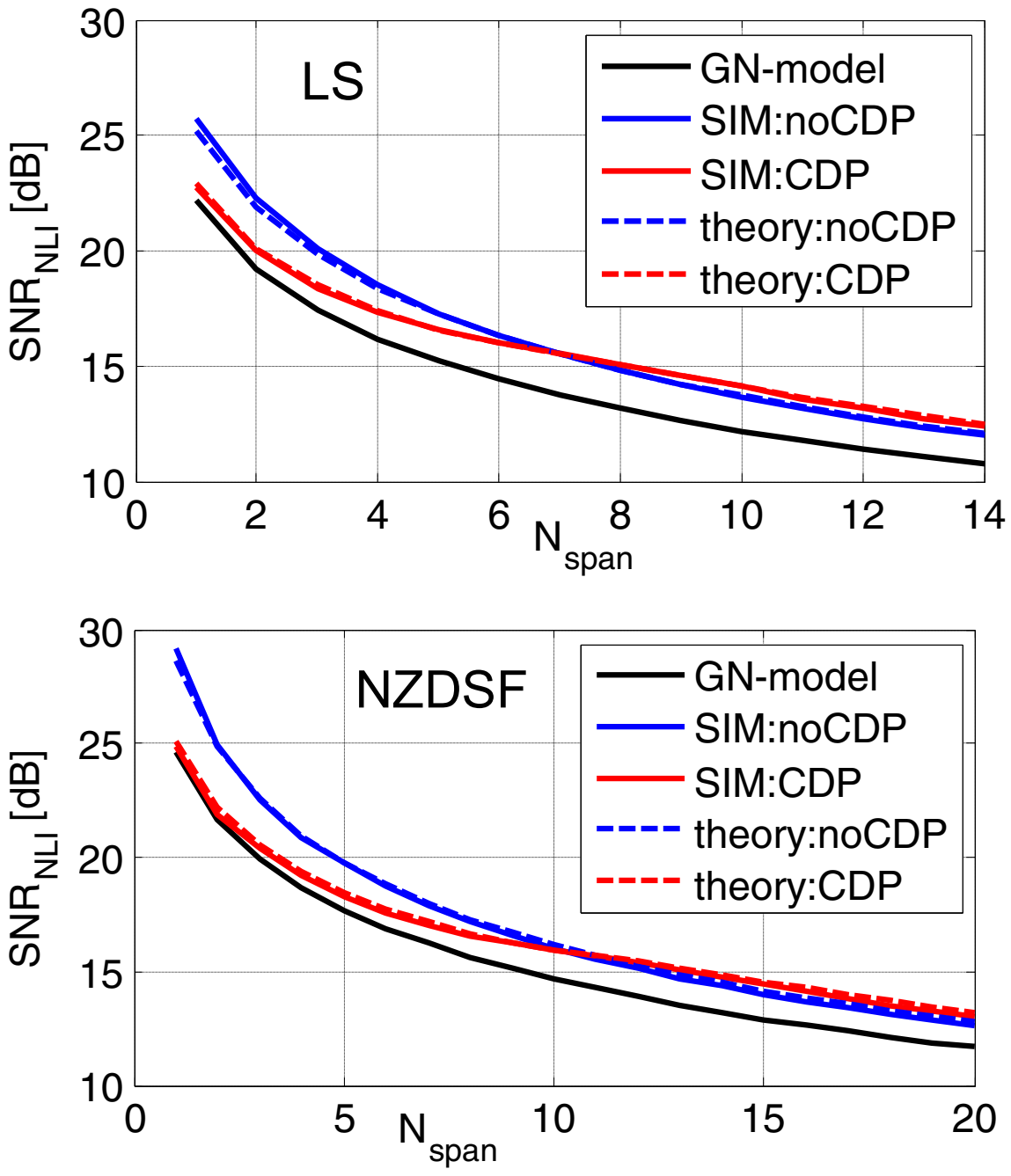

**Fig. 2:** Values of $SNR_{NLI}$ defined by Eq. (1) vs. number of spans, for LS and NZDSF fiber. System is quasi-Nyquist WDM PM-QPSK, 15 channels, 32 GBaud.

A first remark is that simulations and the EGN model match well, which cross-validates the two results. The GN model instead, as expected, overestimates $P_{NLI}$ leading to a lower estimate of $SNR_{NLI}$ than is found by simulation. The blue curves have initially a higher $SNR_{NLI}$ than the red curves but at about mid-link they cross over and, from then on, the red curves have a higher $SNR_{NLI}$. The advantage due to optimum CDP is the gap found at max-reach between the blue and red curves. Its values are rather modest: 0.32, 0.39 and 0.43 dB for SMF, NZDSF and LS, respectively. Note that the gain in $Q^2$ and, approximately, in max-reach is expected to be about 1/3 (dB over dB) of the gain on $SNR_{NLI}$ [7]. This agrees with the 3% max-reach gain found directly as shown in Fig. 1.

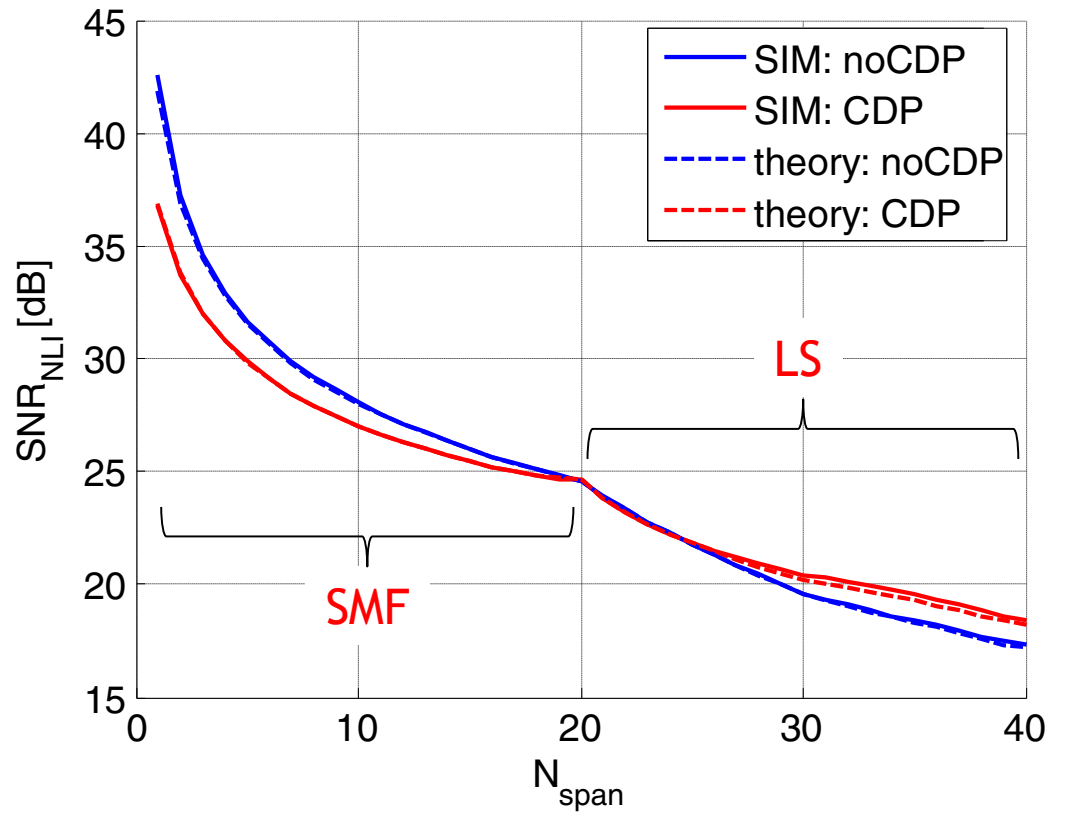

**Fig. 3:** Values of $SNR_{NLI}$ defined by Eq. (1) vs. number of spans, for an inhomogenous system with 20 spans of SMF followed by 20 spans of LS fiber. System is quasi-Nyquist WDM PM-QPSK, 15 channels, 32 GBaud.

## Inhomogeneous fiber link

We then considered an inhomogenous link made up of 20 spans of SMF and 20 spans of LS, with span length 100 km. We considered both the case of SMF followed by LS and vice-versa. Note that, differently from the homogenous case, we fixed the system reach, to ease the calculations burden. The target of the optimization was then to obtain the highest Rx OSNR, defined as:

$$OSNR_{NLI} = P_{RX} /(P_{ASE} + P_{NLI}) \qquad (2)$$

The optimization process involved both $P_{TX,SMF}$ and $P_{TX,LS}$ because the overall NLI noise at the Rx depends on both, in a complex way. As CDP values we assumed either 0 (no CDP) or the $D_{PRE}$ necessary to have zero accumulated dispersion in the middle of the LS section, similar to [6]. This choice of $D_{PRE}$ (-31,600 ps/nm in the SMF-first case and 1,800 ps/nm in the LS-first case) is conceivably close to optimum, but this issue needs further investigation. The optimization was carried out using the EGN model (figure not shown for lack of space) and led to the max OSNR values (over a bandwidth equal to the symbol rate) shown below:

**SMF first:** no CDP, 7.13 dB; CDP, 7.89 dB.

**LS first:** no CDP, 7.70 dB; CDP, 7.94 dB.

Since for the OSNR optimization we used the EGN model exclusively, and even though the model reliability was already tested in the homogenous case, we ran a simulative cross-check of NLI noise prediction, at the optimum launch power. We focused again on $SNR_{NLI}$ of Eq. (1) and we looked at the SMF-first case. The numerator of Eq.(1) must be the optimum launch power into the SMF up to 20 spans and then into the LS up to 40. For CPD: $P_{TX,SMF}$ =0.96, $P_{TX,LS}$=0.5. For no CPD: $P_{TX,SMF}$ =0.96, $P_{TX,LS}$=0.42, all in mW. Fig. 3 shows excellent agreement between EGN-model and simulations, providing a compelling confirmation of the results.

## Discussion

The OSNR gain (coinciding with the $Q^2$ gain) for the inhomogenous link is 0.76 dB in the SMF-first case. It is only 0.24dB in the LS-first case but remarkably the use of CDP equalizes the performance in the two directions. Note that in inhomogenous systems, differently from the homogenous ones, the “1/3 rule” (dB over dB) of OSNR gain vs. $SNR_{NLI}$ gain does not hold, in general. Fig. 3 shows 1dB gain in $SNR_{NLI}$ but the OSNR gain is higher than 1/3 dB (this behavior can be analytically justified). Even though not a game-changer, the potential advantage of CDP appears to be non-negligible, and possibly attractive, for some mixed-fiber scenarios. Further investigation is necessary to explore more mixed-fiber cases and to ascertain that the optimum $D_{PRE}$ is as conjectured here and in [6].

A very important caveat is that in order to impart CDP, a very large $D_{PRE}$ may have to be applied, as in the example of this paper. The PAPR of the Tx signal dramatically increases, especially with PM-QPSK. As reflected in most experimental results from [4]-[6], this may result in substantial penalties, which may obliterate the potential gain. To preserve it, higher-resolution DACs may be needed and the Tx chain must be able to handle the higher PAPR. Also, power consumption is likely to increase significantly, due to the DSP needed for CDP.

This work was supported by CISCO Systems within a SRA contract.

## References

[1] S. J. Savory, "Optimum electronic dispersion compensation strategies for nonlinear transmission, Elect. Lett. **42**, no.7, 2006.

[2] V. Curri et al., “Performance Analysis of Coherent 222-Gb/s NRZ PM-16QAM WDM Systems Over Long-Haul Links” IEEE PTL, 22 (5), 266-268, (2010)

[3] M. S. Alfiad, et al. "A Comparison of Electrical and Optical Dispersion Compensation for 111-Gb/s POLMUX–RZ–DQPSK", JLT, v. 27, no. 16, 2009.

[4] Xiang Liu et al., “406.6-Gb/s PDM-BPSK Superchannel Transmission over 12,800-km TWRS Fiber via Nonlinear Noise Squeezing,” OFC 2013, paper PDP5B.10, (2013)

[5] A. Ghazisaeidi et al., “System Benefits of Digital Dispersion Pre-Compensation for Non-Dispersion-Managed PDM-WDM Transmission” ECOC 2013, paper We.4.D.4, (2013)

[6] Xiang Liu and S. Chandrasekhar, Experimental Study of the Impact of Dispersion Pre-Compensation on PDM-QPSK and PDM-16QAM Performance in Inhomogeneous Fiber Transmission” ECOC 2013, paper P.4.17, (2013)

[7] P. Poggiolini, et al, “The GN model of fiber non-linear propagation and its applications,” JLT **32**, 694-721 (2014).

[8] R. Dar, et al, "Properties of Nonlinear Noise in Long, Dispersion-Uncompensated Fiber Links," Optics Express, vol. 21, no. 22, pp. 25685-25699, Nov. 2013.

[9] A. Carena et al, “On the accuracy of the GN-model and on analytical correction terms to improve it,” arXiv:1401.6946, also submitted to Optics Express.